\documentclass[useAMS,usenatbib]{mn2e}
\usepackage{epsfig,amsmath,amssymb,amsfonts,mathrsfs,latexsym,graphicx,lscape}
\usepackage{hyperref}

\begin{document}

\title[IGR~J17191-2821]{Type I X-ray bursts, burst oscillations and
  kHz quasi-periodic oscillations in the neutron star system
  IGR~J17191--2821}

\author[Altamirano et al.]{D. Altamirano$^1$\thanks{E-mail: d.altamirano@uva.nl}, M. Linares$^1$, A. Patruno$^1$,
  N. Degenaar$^1$, R. Wijnands$^1$, \newauthor M. Klein-Wolt$^1$, M. van der
  Klis$^1$, C. Markwardt$^{2,3}$ \& J. Swank$^{2}$ \\ 
  $^{1}$: Astronomical Institute, ``Anton Pannekoek'', University of
  Amsterdam, Science Park 904, 1098XH, Amsterdam, The Netherlands.\\
  $^2$:   Laboratory for High-Energy Astrophysics, NASA Goddard Space Flight   Center, Greenbelt, MD 20771, U.S.A. \\ 
  $^3$:Department of Astronomy,   University of Maryland, College Park, MD 20742., U.S.A. }

\pagerange{\pageref{firstpage}--\pageref{lastpage}}
\pubyear{2009}

\maketitle

\label{firstpage}

\begin{abstract}

We present a detailed study of the X-ray energy and power spectral
properties of the neutron star transient IGR~J17191-2821. We
discovered four instances of pairs of simultaneous kilohertz
quasi-periodic oscillations (kHz QPOs).  The frequency difference
between these kHz QPOs is between 315~Hz and 362~Hz. We also report on
the detection of five thermonuclear type-I X-ray bursts and the
discovery of burst oscillations at $\sim294$~Hz during three of
them. Finally, we report on a faint and short outburst precursor,
which occurred about two months before the main outburst.  Our results
on the broadband spectral and variability properties allow us to
firmly establish the atoll source nature of IGR~J17191--2821.

\end{abstract}

\section{Introduction}
\label{sec:intro}

Neutron star low--mass X-ray binaries (NS-LMXBs) have been extensively
observed with the \textit{Rossi X-ray Timing Explorer} (RXTE) during
the last 13 years. These observations have led to important
discoveries, such as
persistent and intermittent pulsations in accretion-powered
millisecond X-ray pulsars,
nearly coherent oscillations during X-ray bursts 
and strong quasi-periodic variability on millisecond time-scales (the
so called kilohertz quasi-periodic oscillations; kHz QPOs).

The kHz QPOs are relatively narrow peaks in the power spectrum. They
sometimes occur in pairs and are thought to reflect motion of matter
around the neutron star at the inner edge of the accretion disk
\citep[see, e.g.][]{Miller98}. Up to date, kHz QPOs with similar
characteristics have been detected in about 30 neutron star sources.
While it is often assumed that at least one of the QPOs reflects the
Keplerian motion at the inner edge of the accretion disk \citep[see
  e.g. review by ][]{Vanderklis06}, models have to still
satisfactorily explain the presence and characteristics of both QPOs.

Direct detection of coherent or nearly coherent pulsations is the only
available method to measure the neutron star spin period in LMXBs.
Ten NS-LMXBs out of more than 100 known \citep{Liu07}
have shown coherent millisecond pulsations in their persistent
emission; these systems are known as accretion-powered millisecond
X-ray pulsars \citep[AMXPs; see][for a review of the first six AMXPs
discovered; for the last four see \citealt{Kaaret06},
\citealt{Krimm07}, \citealt{Casella08} and
\citealt{Altamirano08b}]{Wijnands05}.
Sixteen sources to date (including the one presented in this paper)
have shown nearly coherent millisecond oscillations during thermonuclear
Type-I X-ray bursts \citep{Watts08,Watts09}.
As the X-ray burst evolves, the oscillation frequency typically
increases by a few Hz, approaching an asymptotic value ($\nu_{BO}$)
which is stable for a given source from burst to burst. This
asymptotic frequency is thought to trace within a few Hz the spin
frequency ($\nu_s$) of the neutron star \citep{Strohmayer96}.
The AMXPs SAX~J1808--3658 \citep{Chakrabarty03}, XTE~J1814--338
\citep{Strohmayer03a}, Aql X-1 \citep{Casella08} and most recently
HETE~J1900.1--2455 \citep{Watts09} have all shown that
$\nu_s\simeq\nu_{BO}$, strongly supporting the idea that $\nu_{BO}$ is
a good tracer of $\nu_s$.

\begin{figure} 
\centering
\fbox{\resizebox{1\columnwidth}{!}{\rotatebox{0}{\includegraphics{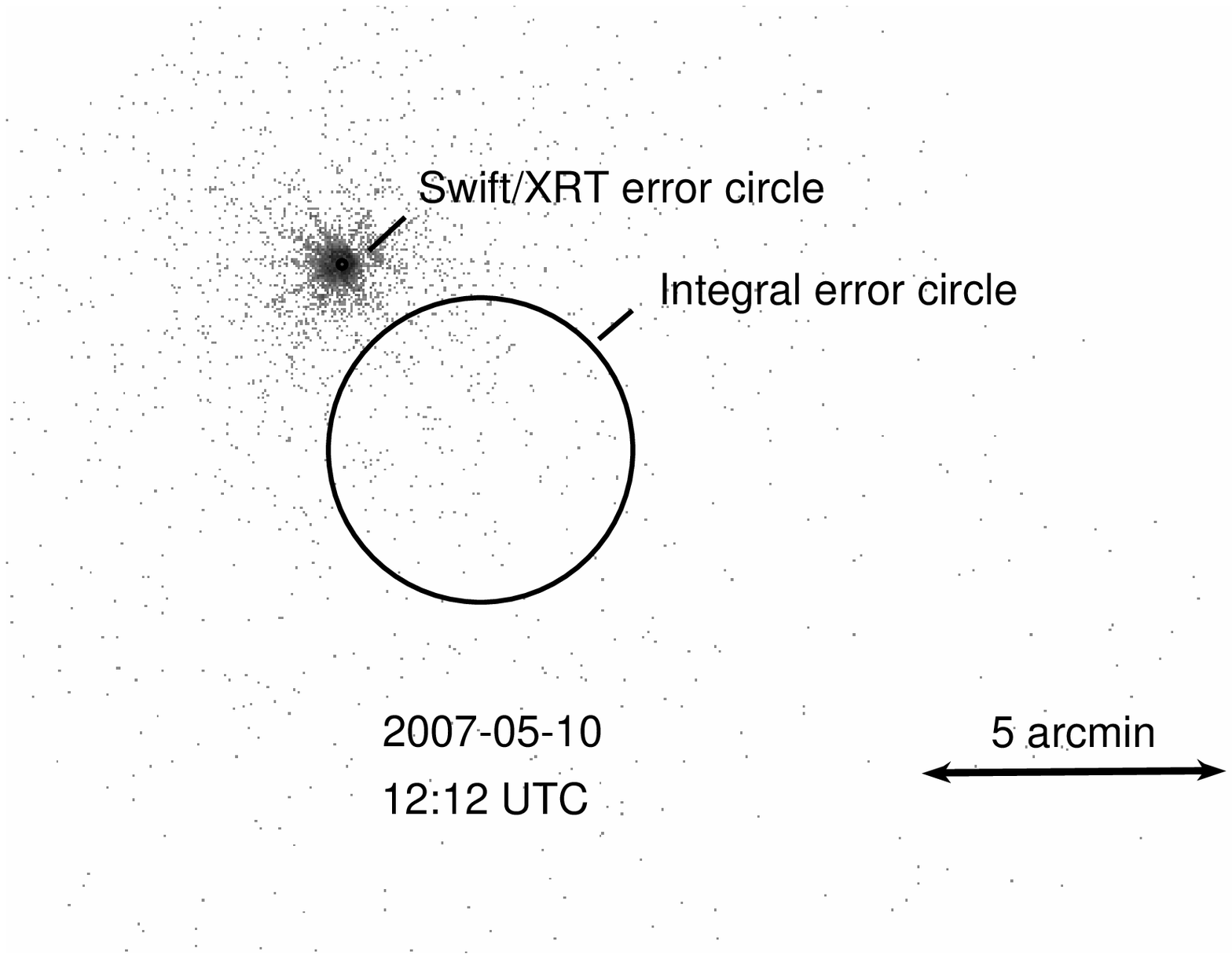}}}}
\caption{Swift/XRT image of IGR~J17191--2821. The black circle
  represents the INTEGRAL error for the source position.  The source
  showed an short (days) outburst precursor or flare on March 8. The
  long outburst studied in this work occurred in May (Swift/XRT
  detection is showed as bright source on the top-left of this
  figure). }
\label{fig:ds9}
\end{figure}

Since the discovery of burst oscillations and two simultaneous kHz
QPOs in neutron star LMXBs, it was suggested that there was a relation
between the neutron star spin frequency, $\nu_s$, and the kHz QPO
frequency difference $\Delta\nu=\nu_u -\nu_{\ell}$.
Although the frequency of the QPOs was found to vary
($150\lesssim\nu_{\ell}\lesssim900$~Hz and $350\lesssim\nu_u
\lesssim1200$~Hz), initial measurements revealed that $\Delta\nu$ was
consistent with being constant\footnote{At least in those sources in
  which $\nu_{BO}$ was known. For example, $\Delta\nu$ in the neutron
  star LMXB Sco~X-1 is known to vary \citep[see,
    e.g.][]{Vanderklis97}. } and equal to the asymptotic burst
oscillation $\nu_{BO}$ \citep[see, e.g., ][]{Strohmayer96}.
This was the main motivation for beat-frequency models such as the
sonic-point model \citep{Miller98}, which proposed that $\nu_u$
reflects the Keplerian frequency at the inner edge of the disk, and
that $\nu_{\ell}$ was the beat between $\nu_u$ and $\nu_s$.
As new observations revealed more sources showing both $\nu_{BO}$ and
twin kHz QPOs, the relation became more complex (see
\citealt{Mendez07} and \citealt{Vanderklis08} for a detailed
historical overview).
As a summary, NS-LMXBs are often classified as \textit{fast} or
\textit{slow} rotators, depending on whether the spin frequency is
higher or lower than $\sim400$~Hz, respectively \citep{Muno01}.
It was found that generally the fast rotators follow
$\Delta\nu\simeq\nu_s/2$, while slow rotators follow
$\Delta\nu\simeq\nu_s$ \citep[see ][ and references
  therein]{Wijnands03,Linares05}. This of course implies that there is
no one-to-one relation between $\Delta\nu$ and $\nu_s$.
Furthermore, it is now known that $\Delta\nu$ is not always consistent
with being constant for sources with known $\nu_{BO}$ \citep[see,
  e.g.,][ and references within]{Mendez98a,Jonker02,Barret06}.

The question of whether $\Delta\nu$ and $\nu_s$ are physically related
is still under debate. While current data might still be compatible
with a bimodal relation, recent results suggest that this might not be
the case. For example, \citet{Mendez07} suggested that $\Delta\nu$ and
$\nu_s$ are unrelated and that the division between fast and slow
rotators may be just an effect of the low number of sources showing
both phenomena \citep[see also][]{Yin07}. Recently,
\citet{Strohmayer08} reported the discovery of burst oscillations at
414.7 Hz in the LMXB 4U~0614+091; if confirmed, these results would
imply a spin frequency that is inconsistent with either
$\Delta\nu\simeq\nu_s/2$ or $\Delta\nu\simeq\nu_s$ relations.

\subsection{IGR~J17191-2821}\label{sec:source}

IGR~J17191-2821 was discovered by the IBIS/ISGRI instrument aboard
INTEGRAL during observations of the galactic bulge monitoring
\citep{Kuulkers07a} conducted between March 2 and 4, 2007
\citep{Turler07}.
The source was detected in the 20--40 and the 40-80 keV bands with
significances of 10.4 and 4$\sigma$, and fluxes of $8.9\pm0.9$ and
$5.8\pm1.4$ mCrab, respectively.
The position of the source was first reported as (RA,DEC) =
(259$^{\circ}$.77, --28$^{\circ}$.35) (J2000) with an accuracy of 2.5
arcmin.

On March 3, 2007, RXTE Galactic bulge scans \citep{Swank01} detected a
source at a position consistent with that reported by \citet{Turler07}
at an intensity of $10\pm1$ mCrab in the 2-10~keV band
\citep{Swank07}. Follow-up RXTE observations performed three to four
days later did not detect the source with a 3$\sigma$ upper limit of
1.2 mCrab. Eight years of Galactic bulge scans were reanalyzed for
contributions from a source at the position of IGR~J17191-2821, but no
flares brighter than 2 mCrab were found.  These non-detections showed
that this source is active relatively infrequently \citep{Swank07}.

During further Galactic bulge scan observations on April 29 and
May 2, 2007, IGR~J17191--2821 was detected again, but at a
level of ~30 and ~70 mCrab (2--10 keV). This suggested that the
previous detection was a flare or outburst precursor. Several
Astronomical Telegrams at this time communicated on the discovery of
Type I X-ray bursts \citep{Klein07a,Markwardt07}, burst oscillations
\citep{Markwardt07} and several episodes of kHz QPOs \citep{Klein07b}.

In this work, we present an intensive study of Type I X-ray burst
characteristics, burst oscillations and kHz QPOs of this newly
discovered neutron star LMXB IGR~J17191-2821.

\section{OBSERVATIONS AND DATA ANALYSIS}
\label{sec:dataanalysis}

\subsection{Light curves, color diagrams and timing
  analysis}\label{sec:lcandccd}

We use data from the Rossi X-ray Timing Explorer (RXTE) Proportional
Counter Array \citep[PCA; for instrument information
see][]{Zhang93,Jahoda06}. To study the long-term (days/months) $L_x$
behavior of the source, we used the PCA monitoring observations of the
galactic bulge \citep{Swank01}. These observations are performed nine
months of the year (as parts of the months of November, December,
January and June are excluded due to solar constraints).  
The accuracy in the position of the PCA bulge scans is about 15
arcmin; the light curves are given in the $\sim2-10$ keV energy band.

To study the short-term (minutes or less) variability, we use PCA
pointed observations. For IGR~J17191--2821 there were 18 observations
in one data set (92052-10) containing $\sim2.5$ to $\sim10$~ksec of
useful data per observation.  We use the 16-s time-resolution Standard
2 mode data to calculate X-ray colors.  
Hard and soft color are defined as the 9.7--16.0 keV / 6.0--9.7 keV
and 3.5--6.0 keV / 2.0--3.5 keV count rate ratio, respectively, and
intensity as the 2.0--16.0 keV count rate.
We removed Type I X-ray bursts from the data as well as corrected by
deadtime effects and for the contribution of the background.  We
normalized colors and intensities by those of the Crab Nebula
\citep[see][ see table 2 in \citealt{Altamirano08} for average colors
  of the Crab Nebula per PCU]{Kuulkers94,Straaten03}.

For the Fourier timing analysis we used the Event mode
E\_125us\_64M\_0\_1s.  Leahy-normalized power spectra were constructed
using data segments of 128 seconds and 1/8192~s time bins such that
the lowest available frequency is $1/128$~Hz and the Nyquist frequency
4096~Hz.  No background or deadtime corrections were performed prior
to the calculation of the power spectra. 
Unless stated explicitly, in our fits we only include those
Lorentzians for which we can measure the integrated power with an
accuracy of at least 3$\sigma$, based on the (negative) error in the
power integrated from 0 to $\infty$. For the kHz QPOs, we report the
centroid frequency $\nu_0$, the full width at half maximum (FWHM) and
the rms amplitude. The quoted errors use $\Delta\chi^2 = 1.0$.  The
upper limits quoted in this paper correspond to a 95\% confidence
level ($\Delta\chi^2=2.7$).

\subsection{Energy spectra of the persistent emission}

For the PCA, we used the Standard 2 data of PCU 2, which was active in
all observations. The background was estimated using PCABACKEST
version 6.0 (see FTOOLS). We calculated the PCU 2 response matrix for
each observation using the FTOOLS routine PCARSP V10.1.
For the HEXTE instrument, spectra were accumulated for cluster B (as
cluster A stopped rocking in October 2006), excluding the damaged
detector and averaging both rocking directions to measure the
background spectrum.
 Dead time corrections of both source and background spectra were
 performed using HXTDEAD V6.0. The response matrices were created
 using HXTRSP V3.1.
Both for PCA and HEXTE, we filtered out data recorded during, and up
to 30 minutes after passage through the South Atlantic Anomaly
(SAA). We only used data when the pointing offset from the source was
less than 0.02 degrees and the elevation of the source respect to the
Earth was greater than 10 degrees. Using XSPEC V11.3.2i
\citep{Arnaud96}, we fitted simultaneously the PCA and HEXTE energy
spectra using the 3.0--25.0~keV and 20.0--200.0~keV energy bands,
respectively. We used a model consisting of a disk blackbody and a
power law, absorbed with an equivalent Hydrogen column density of
0.3$\times$10$^{22}$ cm$^{-2}$ \citep{Klein07c}, which gave a good fit
in all the observations ($\chi^2/dof<1.1$).

\subsection{Type I X-ray bursts}\label{sec:resonbo}

We examined the Standard~1 mode data (2--60 keV, 0.125 seconds time
resolution, no energy resolution) of the 18 observations for Type I
X-ray bursts; we found 5 episodes (see Table~\ref{table:bursts}).
We searched each burst for coherent pulsations using the $Z^2_n$
statistic \citep{Strohmayer99}. We computed $Z^2_1$ (i.e. assuming
that the signal is sinusoidal) throughout the bursts using a sliding 2
seconds window with a step of 0.125 sec. The $Z^2_1$ statistic has the
same statistical properties as a Leahy normalized power spectrum,
which means that for a purely random Poisson process, the powers
follow a $\chi^2$ distribution with 2 degrees of freedom
\citep{Strohmayer99}. We searched the 30--4000~Hz frequency range in
the 2--60 keV band and in narrower energy bands (Sec.~\ref{sec:dis}).

\begin{table*}
\center
\caption{X-ray bursts in IGR~J17191-2821} \label{table:bursts}
\begin{tabular}{ccccccc}
\hline
Number          &  ObsID         &  Start time of the    &  Num. of  &  Flux$^a$             & Osc.         & Osc. rms amplitude$^b$     \\ 
                &                &  burst (UT, 2007)   &  PCUs on  & ($10^{-8}$ erg s$^{-1}$ cm$^{-2}$) & (yes/no)     &(2--17 keV) \\
\hline
 1              & 92052-10-01-00 & May 4 02:32:06     &   2       & $2.5\pm0.3$                     &  yes         &  $6.9\pm0.6 \%$            \\
 2              & 92052-10-05-00 & May 7 02:39:39     &   2       & $2.9\pm0.3$                     &  yes         &  $5.0\pm1.0 \%$            \\
 3              & 92052-10-03-01 & May 7 05:51:12     &   2       & $2.7\pm0.3$                     &  no          &  $<3 \%$            \\
 4              & 92052-10-06-00 & May 8 17:08:53     &   3       & $2.6\pm0.2$                     &  yes         &  $10.2\pm1.5 \%$            \\
 5              & 92052-10-06-01 & May 8 20:38:44     &   2       & $2.1\pm0.3$                     &  no          &  $<3 \%$            \\
\hline
\end{tabular}\\
\flushleft
\vspace{0.2cm}
$^{a}$:Bolometric peak flux.  \\ 
$^{b}$:Integrated amplitude of the oscillations in the 2--17 keV range. \\
\end{table*}

We also created energy spectra every 0.25 sec from the Event mode
(E\_125us\_64M\_0\_1s) data of all the PCUs that were on during the
burst.
Given the high count rates during the peak of the bursts, we corrected
each energy spectrum for dead-time using the methods suggested by the
RXTE team\footnote{http://heasarc.gsfc.nasa.gov/docs/xte/recipes/pca\_deadtime.html}.
For each energy spectrum, we created the corresponding response matrix
using the latest information available on the response of the
instrument at the relevant times.
As is common practice, we used as background the energy spectrum of
the persistent emission taken seconds before each burst.
(We used 100 sec of the persistent emission to calculate the spectrum.
However, we found no significant differences in the fits when the
persistent-emission before or after the burst was used, or when using
data-segments of different lengths -between 100 and 500 seconds-).
We used a black-body model to fit the resulting burst spectra\footnote{
We assumed that the X-ray spectra after the persistent emission
has been subtracted are Planckian and that the observed luminosity of
the source is:

\begin{displaymath}L = 4 \pi \sigma T^4 R^2, \end{displaymath}

\noindent so the unabsorbed bolometric X-ray flux may be determined
using

\begin{displaymath}F_{bol} = \sigma T^4 (R/D)^2, \end{displaymath} 

where $\sigma$ is the Stefan-Boltzmann constant, \textit{T} is the
black-body temperature, \textit{R} the neutron star photosphere
radius, and \textit{D} the distance to the source. The ratio $(R/D)^2$
is the normalization of the black-body model we used
(\textit{bbodyrad} -- see Xspec manual for details).
We note that X-ray burst spectra are generally well described by
black-body emission, however, the emission from the neutron star and
its environment (e.g. accretion disk) is expected to be more complex
than simple black-body emission \citep[see, e.g., ][ and references
  therein]{Vanparadijs82,London84,Kuulkers03}.}.

\section{Results}\label{sec:results}

\subsection{Position of the source}\label{sec:position} 


Swift observed the source twice on March 8, 2007 (at 02:37 UT
for a total of $\sim800$ sec, and at 10:44 UT for a total of
$\sim1600$ sec -- both observations where performed in PC
mode). These are the only Swift observations performed before the
bright outburst (see Section~\ref{sec:lc}).
The observations were taken in Photon Counting mode and did not show a
source within the INTEGRAL error circle with an upper limit
of 0.0021 cnts s$^{-1}$ (at a 95\% confidence level). Assuming a
galactic absorption Nh of $3.4 \times 10^{21}$ cm$^{-2}$ and a photon index
of 1.8, this countrate translates into an upper limit of $\sim8.7
\times10^{-14}$ erg s$^{-1}$ cm$^{-2}$ on the unabsorbed flux
(approximately 0.004 mCrab in the 2-10~keV range).
We found a source in the first Swift observation (i.e. on March 8 at
2:37 UT) located at (RA, DEC) = (259.8114, -28.3005) (J2000, with an
error of $\sim9$ arcsec), i.e., at a distance of about 3.5 arcminutes
from the position of the source as measured by INTEGRAL.
We measured an average count rate of 0.0087 cnts s$^{-1}$, which
corresponds to an unabsorbed flux of $3.6 \times 10^{-13}$ erg s$^{-1}$
cm$^{-2}$ or 0.02 mCrab (2-10 keV, assuming the same Nh and photon
index as above). 
In the second Swift observation this source is not significantly
detected (a total of three photons within a 10 pixel source error
circle and zero from a background region 3 times as large).  We place
a 95\% confidence upper limit on the 2-10 keV unabsorbed flux of
$\sim2.0 \times 10^{-13}$ erg s$^{-1}$ cm$^{-2}$ (employing
PIMMS\footnote{http://heasarc.gsfc.nasa.gov/Tools/w3pimms.html} for an
absorbed power-law spectrum with $N_h = 3.4 \times 10^{21}$ cm$^{-2}$
and a photon index of 1.8, and applying the prescription for low
number statistics given by \citealt{Gehrels86}).
In Figure~\ref{fig:ds9} we show a Swift/XRT image during
IGR~J17191--2821 outburst (May 10th, 2007).

As noted by \citet{Klein07c}, the formal INTEGRAL error circle on the
position of IGR~J17191--2821 would suggest that the faint source we
detected with Swift is unrelated to IGR~J17191--2821 (see
Figure~\ref{fig:ds9}). However, given the systematic uncertainties in
the INTEGRAL position, both sources are probably one and the same. We
note that this is not the first case in which the true position of a
transient laid outside the reported INTEGRAL-IBIS/ISGRI error circle
\citep[see e.g.][]{Kuulkers07}.


\begin{figure} 
\centering
\resizebox{1\columnwidth}{!}{\rotatebox{0}{\includegraphics{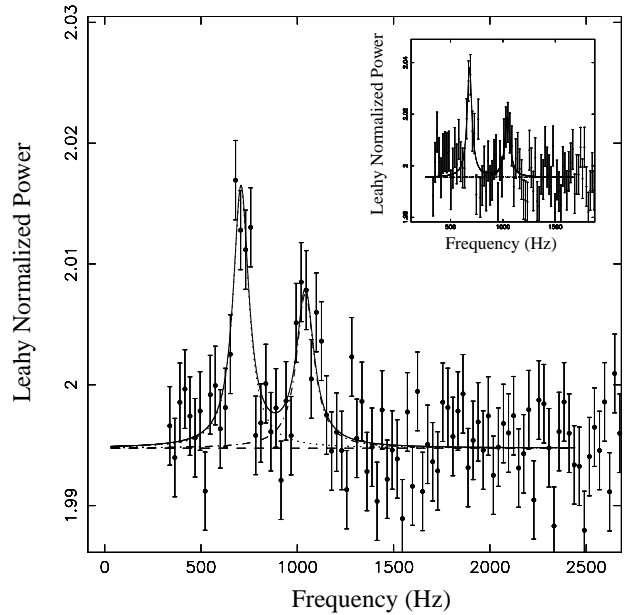}}}
\caption{In the main panel we show the average kHz QPOs of all data
  reported in Table~\ref{table:kHzqpos}, after the shift--and--add
  methods was applied; the lower kHz QPO was shifted to the arbitrary
  frequency of 700~Hz (see Section~\ref{sec:kHz}). Inset: kHz QPOs in
  observation 92052-10-05-00; no shift--and-add was applied.}
\label{fig:kHzqpos}
\end{figure}

\subsection{Thermonuclear X-ray bursts and the distance to the source}\label{sec:dis}

We found 5 Type I X-ray bursts (see Table~\ref{table:bursts}). All
bursts showed similar temperature, radius and bolometric flux profiles
(not shown). The temperature and flux profiles were all single peaked,
and reached their maxima within a second.
The maximum temperature (\textit{kT}) was always between 2 and 3 keV
and the peak bolometric fluxes were in the $2-3 \times 10^{-8}$ erg
s$^{-1}$ cm$^{-2}$ range (Table~\ref{table:bursts}).
In all cases, the black-body radius remained approximately constant
after the peak and was usually constrained between 5 and 10 km
(assuming a distance of 11 kpc).
None of the bursts showed indications of photospheric radius
expansion.  By using the highest measured bolometric peak flux of $3
\times 10^{-8}$ erg s$^{-1}$ cm$^{-2}$ we can estimate an upper limit on
the distance.
We find a distance $D<11$~kpc when using the empirically determined
Eddington luminosity of $3.79 \times 10^{38}$ erg s$^{-1}$ \citep[for
  bursts showing photospheric radius expansion -- ][]{Kuulkers03}.
Using a more complex approximation\footnote{The approximation was
  recently used by \citet{Galloway08} to compare a sample of more than
  a thousand X-ray burst from different sources. The distance is given
  by: \vspace{0.5cm}

\begin{math}
D = 8.6 \cdot 
(\frac{Flux_{Bol}}{3\cdot10^{-8} erg \ cm^{-2} \ s^{-1}})^{-1/2} \cdot
(\frac{M_{NS}}{1.4M_{\odot}})^{1/2} \cdot \\
~~~~~~~~~~~ \cdot (\frac{1+z(R)}{1.31})^{-1/2} \cdot
(1+X)^{-1/2} kpc
\end{math}
\vspace{0.5cm}

where $M_{NS}$ is the mass of the neutron star in solar masses, $X$ is
the mass fraction of hydrogen in the neutron star atmosphere and z(R)
is the term that takes into account the gravitational redshift at the
photosphere \citep[were $1+z(R) = (1-2GM_{NS}/Rc^2)^{-1/2}$, $G$ is
the gravitational constant, $c$ the speed of light and $R$ the radius
measured at the photosphere -- see][]{Galloway08}. }
and standard values for the mass and the radius of the neutron star
(i.e. $M_{NS}=1.4M_{\odot}$ and $R=10$~km), we found $D<8.6$ and
$D<6.6$ for hydrogen mass fractions of $X=0$ and $X=0.7$,
respectively. Larger values of $R$ give higher upper limits.


\subsection{kHz QPOs}\label{sec:kHz}

We searched the averaged power spectrum of each observation for the
presence of significant kHz QPOs at frequencies $\gtrsim200$~Hz.
In each case, we fitted the power spectra between 200 and 4000~Hz with
a model consisting of one or two Lorentzians and a constant to account
for the presence of QPOs and Poisson noise, respectively.
We found that 12 out of the 17 observations show significant QPOs in
the 605-1185 frequency range.
In 4 observations we detected 2 simultaneous kHz QPOs (see
Table~\ref{table:kHzqpos}).
The lower kHz QPO frequency was between 680 and 870~Hz, with single
trial significance between 5.7 and 10$\sigma$.
The upper kHz QPO frequency was between 1037 and 1085~Hz, with single
trial significances between 3.0 and 3.7$\sigma$. $\Delta\nu$ showed no
significant changes and was always consistent with $350$~Hz (see
Table~\ref{table:kHzqpos}).

We tried to better constrain $\Delta\nu$ by using the shift--and--add
method as described by \citet{Mendez98a}.  We first tried to trace the
detected kilohertz QPO using a dynamical power spectrum
\citep[e.g. see figure 2 in ][]{Berger96} to visualize the time
evolution of the QPO frequency, but the signal was too weak to be
detected on timescales shorter than the averaged observation.
Therefore, for each observation we used the fitted averaged frequency
to shift each lower kHz QPO to the arbitrary frequency of 700~Hz.
Next, the shifted, aligned, power spectra were averaged. The average
power spectrum was finally fitted in the range 300--2048~Hz so as to
exclude the edges, which are distorted due to the shifting method.  To
fit the averaged power spectrum, we used a function consisting of two
Lorentzians and a constant to fit the QPO and the Poisson noise,
respectively. In this case, the averaged $\Delta\nu$ is $332\pm16$~Hz.
In Figure~\ref{fig:kHzqpos} we show the shift--and-added power
spectrum and a representative example of the single observation power
spectrum with two kHz QPOs (see inset).

\begin{table*}
  \caption{Observations and kHz QPOs} \label{table:kHzqpos}
\centering
\begin{tabular}{|c|ccc|ccc|ccc|c|}\hline
ObsID          & MJD      &Aver.     &    PCUs   & \multicolumn{3}{c|}{Lower}   & \multicolumn{3}{c|}{Upper} & \\
(92052-10)  & (days)   &Cts/s$^a$ &    on$^b$ & \multicolumn{3}{c|}{kHz QPO} & \multicolumn{3}{c|}{kHz QPO} & $\Delta\nu$ \\
               &          & /PCU2    &             & $\nu$ (Hz) & FWHM (Hz) & rms (\%)   &  $\nu$ (Hz) & FWHM (Hz) & rms (\%)    & (Hz)  \\
\hline
-01-00 & 54224.09 & 182         & 2 & --	        & --	      & --	      & --	       & --	        & --	& --        \\
-02-00 & 54225.07 & 170         & 3-4 & $870\pm1$	& $11\pm1.7$  &  $8.5\pm0.4$  &	$1185\pm50$    & $220\pm94$	& $9.2\pm1.5$ & $315\pm50$	\\
-02-01 & 54225.14 & 188         & 3 & $866\pm3$	& $38\pm6$    &	 $9.2\pm0.6$  & --              & --	        & --     & -- 	\\
-03-00 & 54226.21 & 172         & 1-3 & $881\pm10$	& $132\pm23$  &  $15\pm1$     & --	       & --	        & --     & --	\\
-05-00 & 54227.10 & 142         & 2-3 & $681\pm5$	& $55.2\pm15$ &	 $10.3\pm0.9$       & $1043\pm10$     &$60^{+27}_{-19}$    & $8.4\pm1.1$	  & $362\pm11$     \\
-03-01 & 54227.23 & 148         & 2 & $730\pm1$	& $13.2\pm2.7$ & $9.7\pm0.6$       & $1075\pm12$     &$55^{+32}_{-22}$   & $8.5\pm1.3$	  & $345\pm12$     \\
-04-00 & 54228.01 & 156         & 3 & $891\pm3$	& $37.7\pm7.5$ &  $9.4\pm0.6$ & --	       & --	        & --	& --	\\
-04-01 & 54228.20 & 151         & 2-3 & $838\pm1$	& $25\pm4$	& $9.3\pm0.4$	& --	      & --               & --& --	\\
-06-00 & 54228.70 & 129         & 2-3 & $946\pm20$	& $151^{+66}_{-41}$ & $10.6\pm1.2$ & --	       & --	         & --& --	\\
-06-01 & 54228.90 & 130         & 2-3 & $702\pm3$	& $37\pm7$	  & $10.4\pm0.6$	        & $1037\pm15$	& $88^{+43}_{-27}$	&$8.8\pm1.1$& $335\pm15$	\\
-07-00 & 54229.06 & 105         & 1-2 & $884\pm13$	& $108^{+53}_{-31}$ & $15\pm2$	& --	& --	& --& --	\\
-07-01 & 54229.98 & 72          & 2 & $706\pm10$	& $83\pm25$	& $18\pm2$	& --	& --	& --& --	\\
-05-01 & 54230.04 & 70          & 2 & $682\pm20$	& $172^{+78}_{-55}$ & $22.4\pm2.8$ & --	& --	& --& --	\\
-08-00 & 54231.09 & 46          & 2 & --	& --	& --	& --	& --	& --& --	\\
-08-01 & 54231.75 & 27          & 2-3 & --	& --	& --	& --	& --	& --& --	\\
-09-00 & 54232.01 & 23          & 3 & --	& --	& --	& --	& --	& --& --	\\
-09-01 & 54232.08 & 21          & 3-4 & --	& --	& --	& --	& --	& --& --	\\
\hline
\end{tabular}
\flushleft
$^{a}$: Background and deadtime corrected averaged count rate for PCU2; this PCU was on during all observations.\\
$^b$: In case the number of active PCUs changed, we report the minimum and maximum number.
\end{table*}
\vspace{1cm}

\subsection{Outburst evolution}\label{sec:lc}

\begin{figure}
\centering
\resizebox{1\columnwidth}{!}{\rotatebox{-90}{\includegraphics{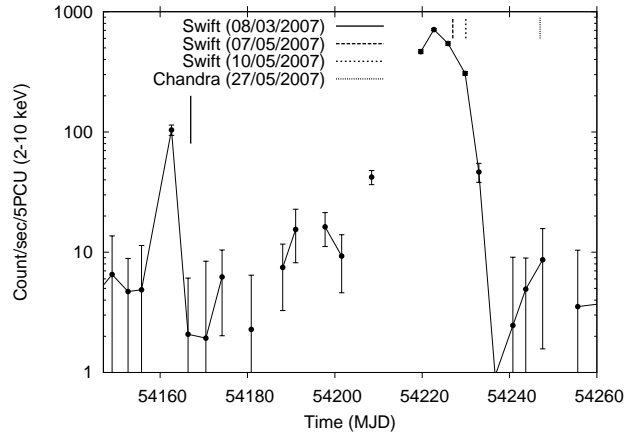}}}
\caption{PCA light curve of IGR~J17191-2821 as seen by the PCA bulge
  scan monitor program \citep{Swank01}. The times of the Swift and
  Chandra observations are marked with vertical lines. Contiguous
  non-zero measurements are connected with lines.}
\label{fig:lc}
\end{figure}

Figure~\ref{fig:lc} shows the PCA light curve of IGR~J17191--2821 as
seen by the PCA bulge scan monitor program \citep{Swank01}. While a
precursor of the outburst (see Section~\ref{sec:source}) occurred at
MJD~54162.6, the full X-ray outburst did not start until 54 days later
(i.e. MJD~54216); it lasted for about 30 days.

On MJD~54247 (May 27th, 2007) the source was not detected anymore with
RXTE, and a Chandra/HRC-I observation was performed.  As reported by
\citet{Chakrabarty07}, the source was not detected in the 1.1 ks
observation within the 30 arcsec of the Swift/XRT position (see
Section~\ref{sec:position}). These authors estimated an upper limit on
the 0.3--10 keV unabsorbed flux of $<8.3 \times 10^{-14}$ erg s$^{-1}$
cm$^{-2}$.

\begin{figure} 
\centering
\resizebox{1\columnwidth}{!}{\rotatebox{0}{\includegraphics{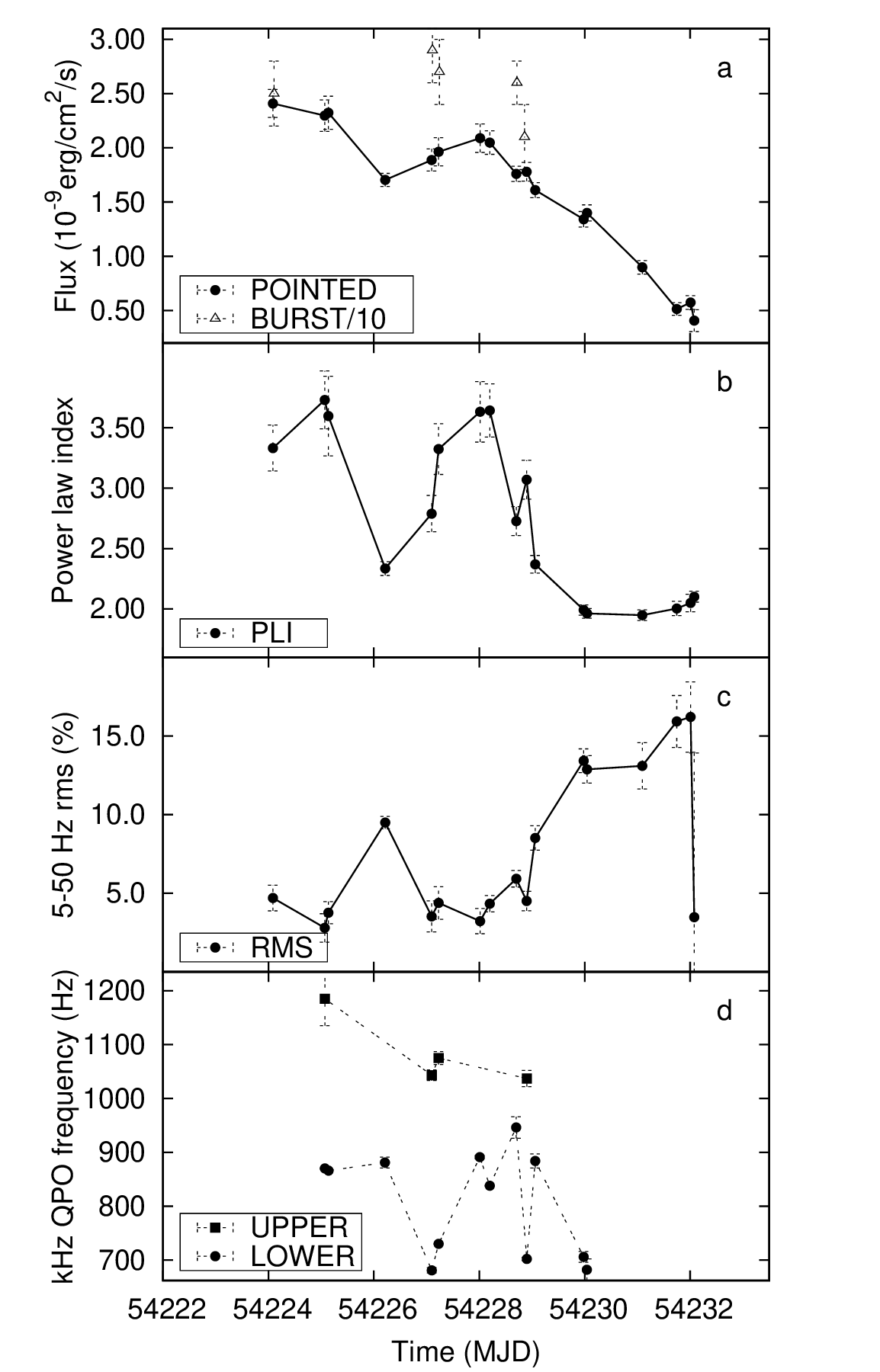}}}
\caption{Spectral and timing properties along the outburst of
  IGR~J17191--2821. From top to bottom, {\it a)}: 2-200~keV unabsorbed
  persistent flux (filled circles) and bolometric peak flux of the
  type I X-ray bursts detected by the PCA (open triangles; peak burst
  flux values are divided by ten for plotting purposes).  {\it b)}:
  power law index obtained from the broad band spectral fits; {\it
    c)}: fractional rms amplitude of the variability between 5 and 50
  Hz in the $\sim$2.5-45~keV energy band and {\it d)}: frequencies of
  the upper (squares) and lower (circles) kHz QPOs detected during the
  outburst.}
\label{fig:manu}
\end{figure}

In Figure~\ref{fig:manu} we show the persistent unabsorbed 2-200~keV
flux (panel $a$), power law index (panel $b$), strength of the broad
band noise (panel $c$) and kHz QPO frequency (panel $d$) as a function
of time during the $\sim$3 weeks of the outburst from which RXTE
pointed observations are available.
The source reached a maximum flux of $\sim$2.5$\times$$10^{-9}$ erg
cm$^2$ s$^{-1}$. Assuming $D<11$~kpc (see Section~\ref{sec:dis}), we
place an upper limit on the outburst peak luminosity of
$4\times$$10^{37}$ erg s$^{-1}$. In panel $a$ of Figure~\ref{fig:manu}
we also plot the (Type I) X-ray burst bolometric peak fluxes at the
time they occurred (as detected by RXTE).

Due to the relatively low count rates collected by the PCA, the
average power spectrum of each observation had low statistical
quality. However, in the brightest (and softest) observations we found
traces of the so-called very low frequency noise (VLFN). 
As a steep power law rising towards low frequencies, this VLFN is a
typical signature of the so called ``banana branch'' (soft state) of
atoll sources \citep[see, e.g., ][for a review]{Vanderklis06}.
When comparing the results showed in the different panels of
Figure~\ref{fig:manu}, we found that the 5-50~Hz averaged fractional
rms amplitude is anti-correlated with the source luminosity, whereas
the frequencies of both kHz QPOs showed no obvious trend. The spectral
index was clearly anti-correlated with the strength of the variability
(i.e, correlates with luminosity). This is similar to what has been
found in other atoll sources \citep[see, e.g., ][for a
  review]{Vanderklis06}, where the strength of the variability and the
spectral index trace the changes in the timing and spectral state
during the outburst.
From Figure~\ref{fig:manu}, we found that the source was initially in
the soft (banana) state and showed a failed transition to the hard
(extreme island) state around MJD 54226. After this, it re-brightened
and returned to the soft state. On MJD 54228 (i.e. two days later) the
luminosity reached a secondary peak and started to decline, while the
source gradually transitioned from the soft (banana) state to the hard
(extreme island) state.  Finally IGR~J17191--2821 faded below the
detection limit of RXTE-PCA.  The timing and spectral properties (and
evolution) allow us to firmly establish the atoll source nature of
IGR~J17191--2821.

\subsection{Burst oscillations}

\begin{figure} 
\centering
\resizebox{1\columnwidth}{!}{\rotatebox{0}{\includegraphics{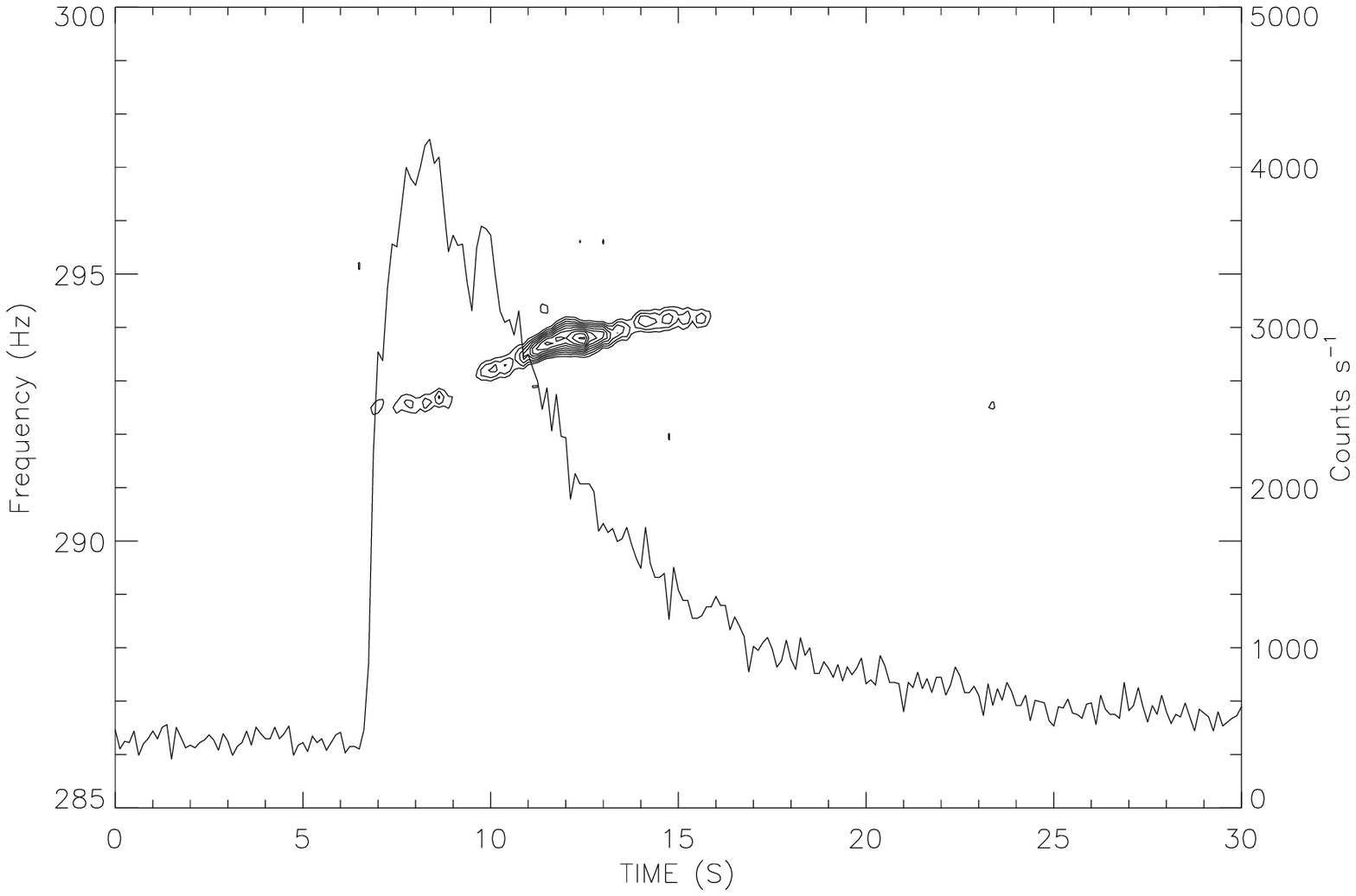}}}
\resizebox{1\columnwidth}{!}{\rotatebox{0}{\includegraphics{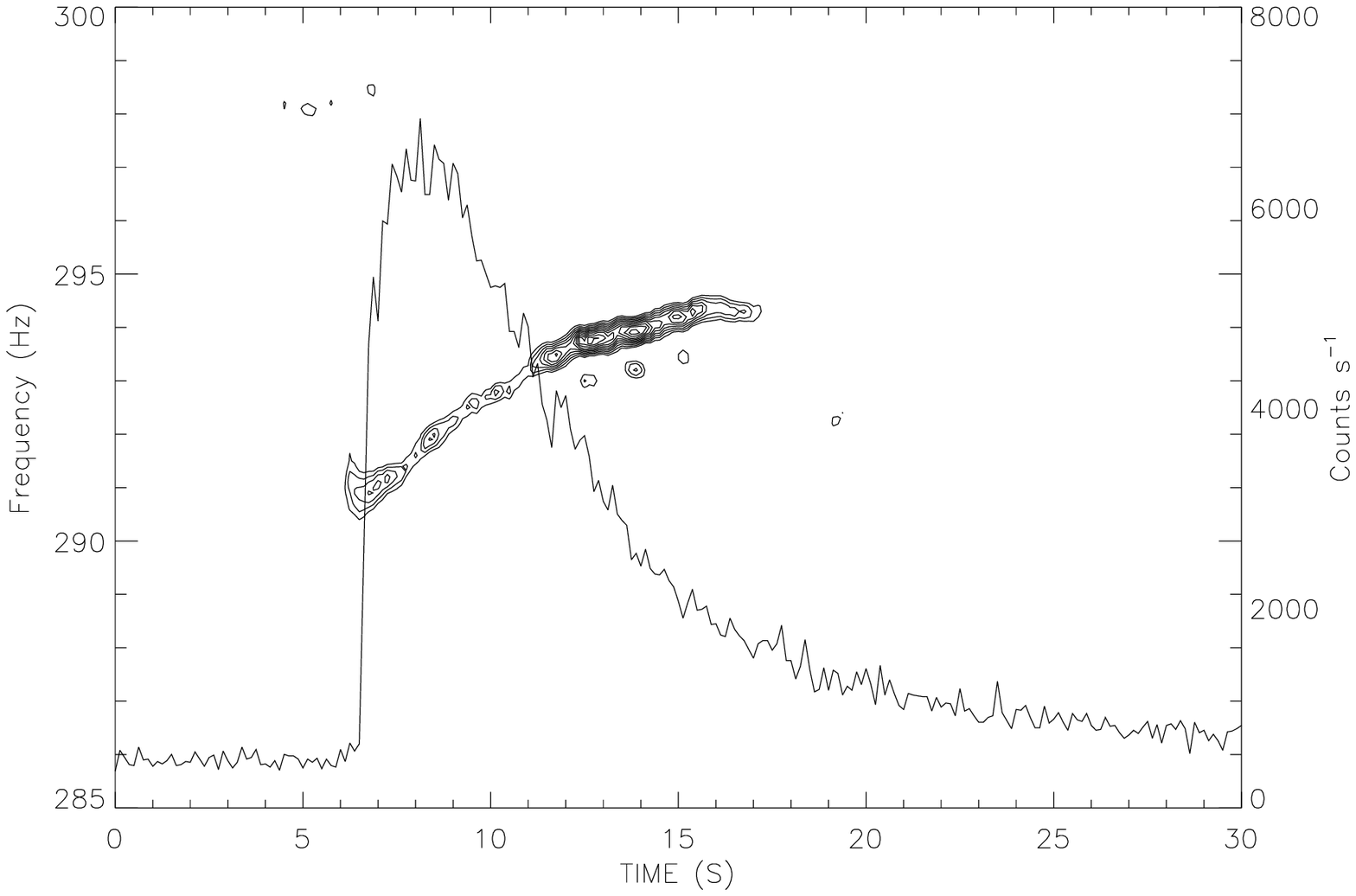}}}
\caption{Lightcurves and dynamical power spectra (2--60 keV) for bursts
  1 and 4 (upper and lower panel, respectively -- see also
  Table~\ref{table:bursts}). The dynamical power spectra were created
  using overlapping 2 s windows, with new windows starting at 0.125 s
  intervals. We used a Nyquist frequency of 4096 Hz. The contours show
  Leahy normalized powers between 20 and 60, increasing in steps of
  5. }
\label{fig:BO}
\end{figure}

By applying the $Z^2_1$ method (see Section~\ref{sec:resonbo}) on the
2--60~keV X-ray burst data, we discovered highly significant
nearly-coherent oscillations in 2 of the 5 X-ray bursts (Burst 1 and 4
in Table~\ref{table:bursts}).
For the remaining 3 bursts, we repeated the $Z^2_1$ analysis using
only data in different energy sub-bands. We found that the
oscillations were significantly detected also in burst number 2, but
only in the 10--25~keV range.  The oscillations are not significantly
detected in either burst 3 or 5. Fractional rms amplitudes (averaged
over the period the signal was significantly detected) and upper
limits are given in Table~\ref{table:bursts}.
In Figure~\ref{fig:BO} we show the dynamical power spectra of bursts 1
and 4 (upper and lower panels, respectively) in the 2--60~keV
range. 
As can be seen, the frequency of the oscillations drifts from
$\sim$291Hz (burst 1) and $\sim292.5$Hz (burst 2), to a frequency
between 294 and 294.5~Hz.
Maximum $Z^2_1$ powers of $\sim63$ were found in both cases.
This type of frequency drifts are typical for burst oscillations
during Type I X-ray bursts \citep[see, e.g., ][ for a
  review]{Strohmayer01}.

\subsection{The energy dependence of the burst oscillations}

\begin{figure} 
\centering
\resizebox{1\columnwidth}{!}{\rotatebox{0}{\includegraphics{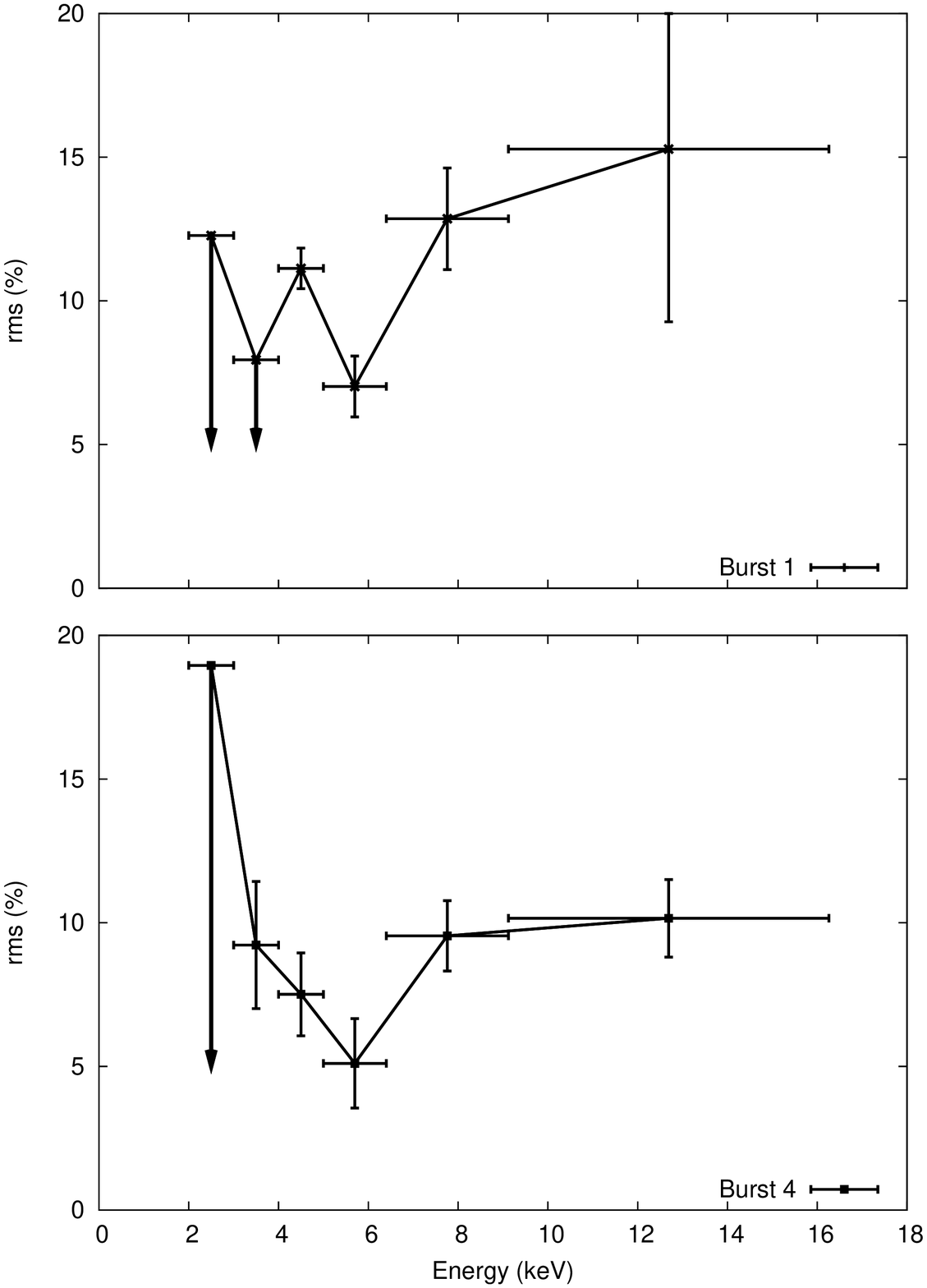}}}
\caption{Fractional rms amplitude versus energy for the oscillations
  in X-ray bursts 1 and 4 (See Table~\ref{table:bursts}). Arrows mark
  upper limits at 95\% confidence level. }
\label{fig:rmsvsenergy}
\end{figure}

We analyzed the energy dependence of bursts 1 and 4 which showed the
strongest oscillations -- see Table~\ref{table:bursts}.
We split the data in six energy bands (from 2.5 keV to 17 keV), and
calculated the amplitude of the pulsations in each band.
To create a pulse profile in each energy band, we selected all the
data with a significant pulse detection in the corresponding power
spectrum, and explored the $P-\dot{P}-\ddot{P}$ space around a given
initial guess value for the pulse period (obtained from the power
spectrum). $\dot{P}$ and $\ddot{P}$ were initially set to zero.
We note that $\dot{P}$ and $\ddot{P}$ do not represent true spin
changes of the neutron star, but comprise all the frequency variations
due, primarily, to the burst oscillation drift seen in the data.
$\dot{P}$ and $\ddot{P}$ are therefore useful to align the phases of
the pulsations (folded in a profile of 32 bins) for each energy band.
We then fitted the pulse profiles with two sinusoids representing the
fundamental and the 1st overtone of the burst oscillations.  The
errors on the fractional amplitudes are calculated using a
$\Delta\chi^{2}=1$. Upper limits are at a 95$\%$ confidence level
(i.e. using $\Delta\chi^{2}=2.7$).
We found no significant second harmonic in any of the energy bands we
chose for either burst, with rms amplitude upper limits in the
$2.5-17$ keV range of 3.7$\%$ and $2.0\%$, for bursts 1 and 4
respectively.
The energy dependence of the burst oscillation rms amplitude is shown
in Figure~\ref{fig:rmsvsenergy}.  The fractional amplitude of the
fundamental is consistent with being constant.
No significant phase lags were detected.

\section{Discussion}\label{sec:discussion}

We present an intensive study of the X-ray variability of the newly
discovered neutron star LMXB IGR~J17191--2821. Our results allow us to
firmly establish the atoll source nature of IGR~J17191--2821. We
detect several episodes of thermonuclear X-ray bursts, some of them
showing burst oscillations that imply a neutron star with spin
frequency of $\sim294$~Hz. We also detect several instances of kHz
QPOs; when two were detected simultaneously, the difference in
frequency is consistent with being constant (See
Table~\ref{table:kHzqpos}).
The energy and broadband power spectra of IGR~J17191--2821 evolved in
a manner consistent with that seen in other neutron star LMXBs: it is
soft when the flux is high and hard when the flux is low. Near the end
of the outburst, IGR~J17191--2821 shows the hardest spectra and
strongest variability, with an rms amplitude above 15\%.
%
%
%

Two months before the main outburst, the source exhibited a very brief
(only days) event which was nearly an order or magnitude less
luminous. It is unclear what the relation of this event is with
respect to the main outburst. However, we note that similar precursors
have been seen before \citep[see, e.g.,][]{Degenaar09}. To our
knowledge, no systematic search has been performed in order to
quantify how common these precursors are and how they can be explained
in the commonly used disk-instability models proposed for outbursts of
X-ray binaries \citep[see][for a review]{Lasota01}.

\begin{figure}
\centering
\resizebox{1.\columnwidth}{!}{\rotatebox{-90}{\includegraphics{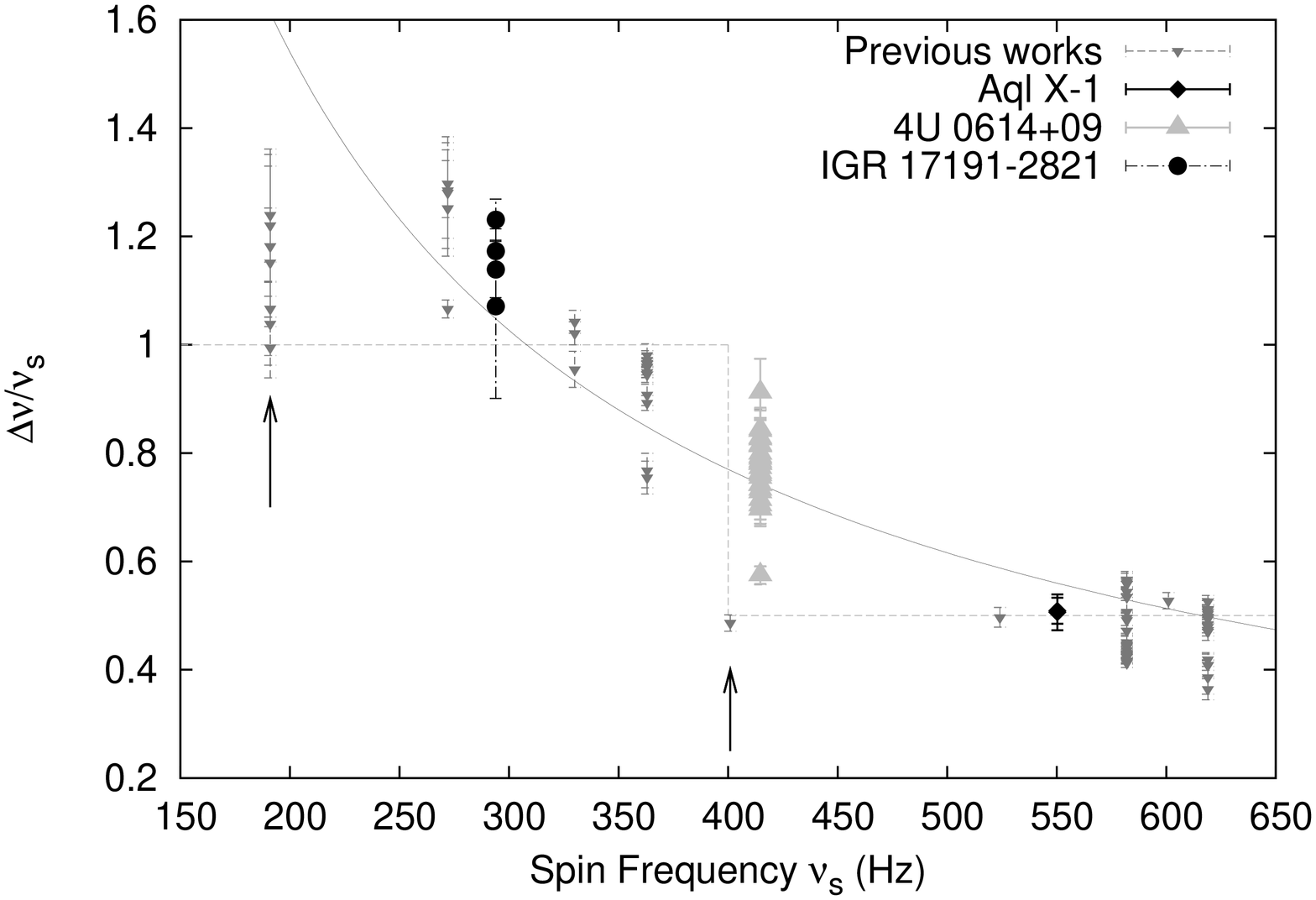}}}
\resizebox{1.\columnwidth}{!}{\rotatebox{-90}{\includegraphics{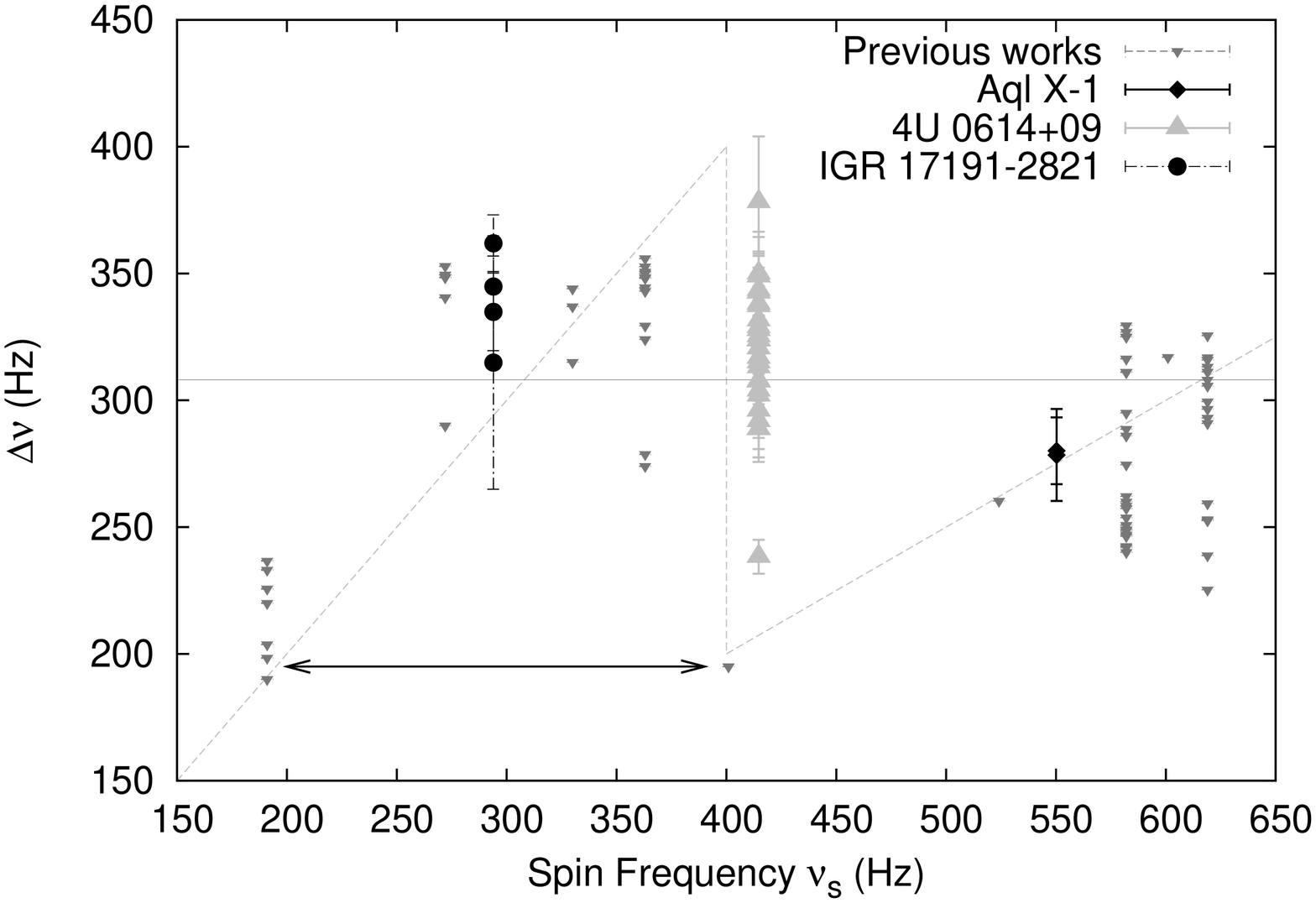}}}
\caption{In the upper panel we plot the ratio between the individual
  measurements of $\Delta\nu=\nu_u-\nu_{\ell}$ and the spin frequency
  $\nu_s$ as a function of $\nu_s$
  \citep[cf.][]{Vanderklis06,Mendez07,Vanderklis08}. Spin frequencies
  are estimated from the persistent pulsations observed in AMXPs
  (marked with the arrows) or from burst oscillations in the
  non-pulsating sources.  
  The lower panel shows the same data as above, but we plot
  $\Delta\nu$ vs. $\nu_s$.
  The dashed lines show the step function $S(\nu_s)=1$ for
  $\nu_s\leq400$~Hz; $S(\nu_s)=0.5$ for $\nu_s>400$~Hz.
  The continuous lines correspond to a constant $\Delta\nu = 308$~Hz
  \citep{Mendez07}.
  For 4U~0614+09 we used the tentative burst oscillation frequency of
  414.7~Hz \citep{Strohmayer08} and the kHz QPOs measurements reported
  by \citet{Straaten00}. For Aql X-1 we used $\nu_s=550$~Hz
  \citep[e.g. ][and references within]{Casella08} and the tentative
  $\Delta\nu$ measurements reported by \citet{Barret08}.
  IGR~J17191--2821 data are from this work. }

\label{fig:deltanu}
\end{figure}

There are currently 12 LMXBs with reported $\Delta\nu$ measurements
whose spin can be estimated from either pulsations in their persistent
X-ray emission or from burst oscillations \citep[][for a recent
  overview]{Vanderklis08}.
In Figure~\ref{fig:deltanu} (upper panel) we plot $\Delta\nu/\nu_s$
vs. $\nu_s$ for all these sources
\citep[cf.][]{Vanderklis06,Mendez07,Vanderklis08}. The dashed line
represents a step function: $\Delta\nu/\nu_s \sim1$ for the slow
rotators ($<400$~Hz) and $\Delta\nu/\nu_s \sim0.5$ for the fast
rotators ($>400$~Hz). Although most of the data seem to be consistent
with this scheme (and our data of IGR~J17191-2821 are as well),
certainly there are points that do not follow this relation.
A clear example is given by the neutron star 4U~0614+09. A tentative
$\sim414.7$~Hz burst oscillation frequency was recently reported for
this source \citep[][ note that we quote this value as tentative since
  it has been detected only once, the signal showed no frequency drift
  as expected from burst oscillations, and it was the first and only
  detection to date of burst oscillations with the Burst Alert
  Telescope on board the Swift telescope]{Strohmayer08}.
As shown in Figure~\ref{fig:deltanu} this burst oscillation frequency
is very close to the discontinuity of the step function
(although given the present data, the discontinuity of the step
function could be anywhere between $\sim360$ and 401 Hz; furthermore,
to our knowledge none of the models predict the exact value at which
$\Delta\nu / nu_s$ should switch from 1 to 0.5).
At the same time, $\Delta\nu/\nu_s$ seems to cover almost the complete
0.5--1. range \citep{Straaten00}, although we note that
93\% of the $\Delta\nu/\nu_s$ measurements are in the 0.69--0.85 range
and the average $\Delta\nu/\nu_s$ using all measurements is
$0.767\pm0.006$ \citep[see also, ][]{Boutelier09}.



\citet{Mendez07} have recently suggested that $\Delta\nu$ and $\nu_s$
are unrelated and that the division between fast and slow rotators may
be just an effect of the low number of sources showing both phenomena
\citep[see also][]{Yin07}. In the lower panel of
Figure~\ref{fig:deltanu} we show $\Delta\nu$ and $\nu_s$ for the same
data plotted in the upper panel. The dashed line in this panel
corresponds to the average $\Delta\nu=308$~Hz \citet{Mendez07}. As can
be seen, the $\Delta\nu$ range of most sources overlaps with this
constant value, except for the two AMXPs SAX~J1808.4--3658 and
XTE~J1807--294, for which $\Delta\nu$ falls clearly below 300 Hz. The
discrepancy is solved, if the data for these two sources are
multiplied by a factor of 1.5. This was first suggested by
\citet{Mendez07} based on the works of \citet{Straaten05} and
\citet{Linares05}
\footnote{Previous works have shown that the frequencies of the
  variability components observed in atoll sources follow a universal
  scheme of correlations when plotted versus $\nu_u$ \citep[see
    e.g.][and references therein]{Straaten02,Straaten03, Reig04,
    Altamirano08}.
The two AMXPs SAX~J1808.4--3658 and XTE~J1807--294 show similar
relations. However, the relations for these two sources are shifted
with respect to those of the other sources \citep{Straaten05,
  Linares05}. This shift is between the frequencies of the
low-frequency components and $\nu_u$ by a factor around 1.5 (1.45 and
1.59 for SAX~J1808.4--3658 and XTE~J1807--294, respectively)
 and between $\nu_{\ell}$ and $\nu_u$ by a similar factor.  }.
%
We note that not all AMXPs are affected by the same multiplicative
factor \citep[][]{Straaten05}, and furthermore, that the factor might
be independent of whether the neutron star pulsates or not
\citep{Altamirano05}.

Clearly, the present data are not enough to draw any final conclusion
on the relation between $\Delta\nu$ and $\nu_s$; the detection of both
the spin frequency and $\Delta\nu$ in other sources is necessary. The
new instrument ASTROSAT (an Indian multiwavelength Astronomy
Satellite), which is planned to be launched in 2010, will play a major
role in solving this issue as it is likely to increase the sample of
sources with both spin frequency and kHz QPOs measurements.

\textbf{Acknowledgments:} This research has made use of data obtained
from the High Energy Astrophysics Science Archive Research Center
(HEASARC), provided by NASA's Goddard Space Flight Center. We are
grateful to the referee for her/his comments that helped to
strengthen some of the points presented in this paper.


\begin{thebibliography}{57}
\expandafter\ifx\csname natexlab\endcsname\relax\def\natexlab#1{#1}\fi
\expandafter\ifx\csname url\endcsname\relax
  \def\url#1{{\tt #1}}\fi
\expandafter\ifx\csname urlprefix\endcsname\relax\def\urlprefix{URL }\fi

\bibitem[{{Altamirano} et~al.(2005){Altamirano}, {van der Klis}, {M{\'e}ndez}
  et~al.}]{Altamirano05}
{Altamirano} D., {van der Klis} M., {M{\'e}ndez} M., et~al., 2005, \apj, 633,
  358

\bibitem[{{Altamirano} et~al.(2008{\natexlab{a}}){Altamirano}, {Casella},
  {Patruno}, {Wijnands}, \& {van der Klis}}]{Altamirano08b}
{Altamirano} D., {Casella} P., {Patruno} A., {Wijnands} R., {van der Klis} M.,
  Feb. 2008{\natexlab{a}}, \apjl, 674, L45

\bibitem[{{Altamirano} et~al.(2008{\natexlab{b}}){Altamirano}, {van der Klis},
  {M{\'e}ndez} et~al.}]{Altamirano08}
{Altamirano} D., {van der Klis} M., {M{\'e}ndez} M., et~al., Sep.
  2008{\natexlab{b}}, \apj, 685, 436

\bibitem[{{Arnaud}(1996)}]{Arnaud96}
{Arnaud} K.A., 1996, In: {Jacoby} G.H., {Barnes} J. (eds.) Astronomical Data
  Analysis Software and Systems V, vol. 101 of Astronomical Society of the
  Pacific Conference Series, 17--+

\bibitem[{{Barret} et~al.(2006){Barret}, {Olive}, \& {Miller}}]{Barret06}
{Barret} D., {Olive} J.F., {Miller} M.C., 2006, \mnras, 370, 1140

\bibitem[{{Barret} et~al.(2008){Barret}, {Boutelier}, \& {Miller}}]{Barret08}
{Barret} D., {Boutelier} M., {Miller} M.C., Mar. 2008, \mnras, 384, 1519

\bibitem[{{Berger} et~al.(1996){Berger}, {van der Klis}, {van Paradijs}
  et~al.}]{Berger96}
{Berger} M., {van der Klis} M., {van Paradijs} J., et~al., 1996, \apjl, 469,
  L13+

\bibitem[{{Boutelier} et~al.(2009){Boutelier}, {Barret}, \&
  {Miller}}]{Boutelier09}
{Boutelier} M., {Barret} D., {Miller} M.C., Jul. 2009, ArXiv e-prints

\bibitem[{{Casella} et~al.(2008){Casella}, {Altamirano}, {Patruno}, {Wijnands},
  \& {van der Klis}}]{Casella08}
{Casella} P., {Altamirano} D., {Patruno} A., {Wijnands} R., {van der Klis} M.,
  Feb. 2008, \apjl, 674, L41

\bibitem[{{Chakrabarty} et~al.(2003){Chakrabarty}, {Morgan}, {Wijnands}
  et~al.}]{Chakrabarty03}
{Chakrabarty} D., {Morgan} E.H., {Wijnands} R., et~al., Mar. 2003, In: Bulletin
  of the American Astronomical Society, vol.~35 of Bulletin of the American
  Astronomical Society, 657--+

\bibitem[{{Chakrabarty} et~al.(2007){Chakrabarty}, {Krauss}, {Jonker}, {Juett},
  \& {Markwardt}}]{Chakrabarty07}
{Chakrabarty} D., {Krauss} M.I., {Jonker} P.G., {Juett} A.M., {Markwardt} C.B.,
  Jun. 2007, The Astronomer's Telegram, 1096, 1

\bibitem[{{Degenaar} \& {Wijnands}(2009)}]{Degenaar09}
{Degenaar} N., {Wijnands} R., Feb. 2009, \aap, 495, 547

\bibitem[{{Galloway} et~al.(2008){Galloway}, {Muno}, {Hartman}, {Psaltis}, \&
  {Chakrabarty}}]{Galloway08}
{Galloway} D.K., {Muno} M.P., {Hartman} J.M., {Psaltis} D., {Chakrabarty} D.,
  Dec. 2008, \apjs, 179, 360

\bibitem[{{Gehrels}(1986)}]{Gehrels86}
{Gehrels} N., Apr. 1986, \apj, 303, 336

\bibitem[{{Jahoda} et~al.(2006){Jahoda}, {Markwardt}, {Radeva}
  et~al.}]{Jahoda06}
{Jahoda} K., {Markwardt} C.B., {Radeva} Y., et~al., 2006, \apjs, 163, 401

\bibitem[{{Jonker} et~al.(2002){Jonker}, {M{\' e}ndez}, \& {van der
  Klis}}]{Jonker02}
{Jonker} P.G., {M{\' e}ndez} M., {van der Klis} M., 2002, \mnras, 336, L1

\bibitem[{{Kaaret} et~al.(2006){Kaaret}, {Morgan}, {Vanderspek}, \&
  {Tomsick}}]{Kaaret06}
{Kaaret} P., {Morgan} E.H., {Vanderspek} R., {Tomsick} J.A., 2006, \apj, 638,
  963

\bibitem[{{Klein-Wolt} et~al.(2007{\natexlab{a}}){Klein-Wolt}, {Wijnands},
  {Markwardt}, \& {Swank}}]{Klein07c}
{Klein-Wolt} M., {Wijnands} R., {Markwardt} C.B., {Swank} J.H., Mar.
  2007{\natexlab{a}}, The Astronomer's Telegram, 1025, 1

\bibitem[{{Klein-Wolt} et~al.(2007{\natexlab{b}}){Klein-Wolt}, {Wijnands},
  {Swank}, \& {Markwardt}}]{Klein07a}
{Klein-Wolt} M., {Wijnands} R., {Swank} J.H., {Markwardt} C.B., May
  2007{\natexlab{b}}, The Astronomer's Telegram, 1065, 1

\bibitem[{{Klein-Wolt} et~al.(2007{\natexlab{c}}){Klein-Wolt}, {Wijnands},
  {Swank}, \& {Markwardt}}]{Klein07b}
{Klein-Wolt} M., {Wijnands} R., {Swank} J.H., {Markwardt} C.B., May
  2007{\natexlab{c}}, The Astronomer's Telegram, 1075, 1

\bibitem[{{Krimm} et~al.(2007){Krimm}, {Markwardt}, {Deloye} et~al.}]{Krimm07}
{Krimm} H.A., {Markwardt} C.B., {Deloye} C.J., et~al., Oct. 2007, \apjl, 668,
  L147

\bibitem[{{Kuulkers} et~al.(1994){Kuulkers}, {van der Klis}, {Oosterbroek}
  et~al.}]{Kuulkers94}
{Kuulkers} E., {van der Klis} M., {Oosterbroek} T., et~al., 1994, \aap, 289,
  795

\bibitem[{{Kuulkers} et~al.(2003){Kuulkers}, {den Hartog}, {in't Zand}
  et~al.}]{Kuulkers03}
{Kuulkers} E., {den Hartog} P.R., {in't Zand} J.J.M., et~al., 2003, \aap, 399,
  663

\bibitem[{{Kuulkers} et~al.(2007{\natexlab{a}}){Kuulkers}, {Shaw}, {Chenevez}
  et~al.}]{Kuulkers07}
{Kuulkers} E., {Shaw} S., {Chenevez} J., et~al., Feb. 2007{\natexlab{a}}, The
  Astronomer's Telegram, 1008, 1

\bibitem[{{Kuulkers} et~al.(2007{\natexlab{b}}){Kuulkers}, {Shaw}, {Paizis}
  et~al.}]{Kuulkers07a}
{Kuulkers} E., {Shaw} S.E., {Paizis} A., et~al., May 2007{\natexlab{b}}, \aap,
  466, 595

\bibitem[{{Lasota}(2001)}]{Lasota01}
{Lasota} J.P., Jun. 2001, New Astronomy Review, 45, 449

\bibitem[{{Linares} et~al.(2005){Linares}, {van der Klis}, {Altamirano}, \&
  {Markwardt}}]{Linares05}
{Linares} M., {van der Klis} M., {Altamirano} D., {Markwardt} C.B., Dec. 2005,
  \apj, 634, 1250

\bibitem[{{Liu} et~al.(2007){Liu}, {van Paradijs}, \& {van den Heuvel}}]{Liu07}
{Liu} Q.Z., {van Paradijs} J., {van den Heuvel} E.P.J., Jul. 2007, \aap, 469,
  807

\bibitem[{{London} et~al.(1984){London}, {Howard}, \& {Taam}}]{London84}
{London} R.A., {Howard} W.M., {Taam} R.E., Dec. 1984, \apjl, 287, L27

\bibitem[{{Markwardt} et~al.(2007){Markwardt}, {Klein-Wolt}, {Swank}, \&
  {Wijnands}}]{Markwardt07}
{Markwardt} C.B., {Klein-Wolt} M., {Swank} J.H., {Wijnands} R., May 2007, The
  Astronomer's Telegram, 1068, 1

\bibitem[{{M{\'e}ndez} \& {Belloni}(2007)}]{Mendez07}
{M{\'e}ndez} M., {Belloni} T., Oct. 2007, \mnras, 381, 790

\bibitem[{{M{\'e}ndez} et~al.(1998){M{\'e}ndez}, {van der Klis}, {Wijnands}
  et~al.}]{Mendez98a}
{M{\'e}ndez} M., {van der Klis} M., {Wijnands} R., et~al., 1998, \apjl, 505,
  L23+

\bibitem[{{Miller} et~al.(1998){Miller}, {Lamb}, \& {Psaltis}}]{Miller98}
{Miller} M.C., {Lamb} F.K., {Psaltis} D., 1998, \apj, 508, 791

\bibitem[{{Muno} et~al.(2001){Muno}, {Chakrabarty}, {Galloway}, \&
  {Savov}}]{Muno01}
{Muno} M.P., {Chakrabarty} D., {Galloway} D.K., {Savov} P., Jun. 2001, \apjl,
  553, L157

\bibitem[{{Reig} et~al.(2004){Reig}, {van Straaten}, \& {van der
  Klis}}]{Reig04}
{Reig} P., {van Straaten} S., {van der Klis} M., 2004, \apj, 602, 918

\bibitem[{{Strohmayer}(2001)}]{Strohmayer01}
{Strohmayer} T.E., 2001, Advances in Space Research, 28, 511

\bibitem[{{Strohmayer} \& {Markwardt}(1999)}]{Strohmayer99}
{Strohmayer} T.E., {Markwardt} C.B., May 1999, \apjl, 516, L81

\bibitem[{{Strohmayer} et~al.(1996){Strohmayer}, {Zhang}, {Swank}
  et~al.}]{Strohmayer96}
{Strohmayer} T.E., {Zhang} W., {Swank} J.H., et~al., 1996, \apjl, 469, L9+

\bibitem[{{Strohmayer} et~al.(2003){Strohmayer}, {Markwardt}, {Swank}, \& {in't
  Zand}}]{Strohmayer03a}
{Strohmayer} T.E., {Markwardt} C.B., {Swank} J.H., {in't Zand} J., 2003, \apjl,
  596, L67

\bibitem[{{Strohmayer} et~al.(2008){Strohmayer}, {Markwardt}, \&
  {Kuulkers}}]{Strohmayer08}
{Strohmayer} T.E., {Markwardt} C.B., {Kuulkers} E., Jan. 2008, \apjl, 672, L37

\bibitem[{{Swank} \& {Markwardt}(2001)}]{Swank01}
{Swank} J., {Markwardt} K., 2001, in ASP Conf. Ser. 251, New Century of X-ray
  Astronomy, eds. H. Inoue \& H. Kunieda (San Francisco: ASP), 94

\bibitem[{{Swank} et~al.(2007){Swank}, {Markwardt}, {Klein-Wolt}, \&
  {Wijnands}}]{Swank07}
{Swank} J.H., {Markwardt} C.B., {Klein-Wolt} M., {Wijnands} R., Mar. 2007, The
  Astronomer's Telegram, 1022, 1

\bibitem[{{Turler} et~al.(2007){Turler}, {Balman}, {Bazzano} et~al.}]{Turler07}
{Turler} M., {Balman} S., {Bazzano} A., et~al., Mar. 2007, The Astronomer's
  Telegram, 1021, 1

\bibitem[{{van der Klis}(2006)}]{Vanderklis06}
{van der Klis} M., 2006, in Compact Stellar X-Ray Sources, ed. W. H. G. Lewin
  \& M. van der Klis (Cambridge: Cambridge Univ. Press)

\bibitem[{{van der Klis}(2008)}]{Vanderklis08}
{van der Klis} M., Oct. 2008, In: {Wijnands} R., {Altamirano} D., {Soleri} P.,
  et~al. (eds.) American Institute of Physics Conference Series, vol. 1068 of
  American Institute of Physics Conference Series, 163--173

\bibitem[{{van der Klis} et~al.(1997){van der Klis}, {Wijnands}, {Horne}, \&
  {Chen}}]{Vanderklis97}
{van der Klis} M., {Wijnands} R.A.D., {Horne} K., {Chen} W., 1997, \apjl, L97+

\bibitem[{{van Paradijs}(1982)}]{Vanparadijs82}
{van Paradijs} J., Mar. 1982, \aap, 107, 51

\bibitem[{{van Straaten} et~al.(2000){van Straaten}, {Ford}, {van der Klis},
  {M{\'e}ndez}, \& {Kaaret}}]{Straaten00}
{van Straaten} S., {Ford} E.C., {van der Klis} M., {M{\'e}ndez} M., {Kaaret}
  P., 2000, \apj, 540, 1049

\bibitem[{{van Straaten} et~al.(2002){van Straaten}, {van der Klis}, {di
  Salvo}, \& {Belloni}}]{Straaten02}
{van Straaten} S., {van der Klis} M., {di Salvo} T., {Belloni} T., 2002, \apj,
  568, 912

\bibitem[{{van Straaten} et~al.(2003){van Straaten}, {van der Klis}, \& {M{\'
  e}ndez}}]{Straaten03}
{van Straaten} S., {van der Klis} M., {M{\' e}ndez} M., 2003, \apj, 596, 1155

\bibitem[{{van Straaten} et~al.(2005){van Straaten}, {van der Klis}, \&
  {Wijnands}}]{Straaten05}
{van Straaten} S., {van der Klis} M., {Wijnands} R., 2005, \apj, 619, 455

\bibitem[{{Watts} et~al.(2008){Watts}, {Krishnan}, {Bildsten}, \&
  {Schutz}}]{Watts08}
{Watts} A.L., {Krishnan} B., {Bildsten} L., {Schutz} B.F., Sep. 2008, \mnras,
  389, 839

\bibitem[{{Watts} et~al.(2009){Watts}, {Altamirano}, {Linares}
  et~al.}]{Watts09}
{Watts} A.L., {Altamirano} D., {Linares} M., et~al., Apr. 2009, ArXiv e-prints

\bibitem[{{Wijnands}(2005)}]{Wijnands05}
{Wijnands} R., 2005, ArXiv Astrophysics, astro-ph/0501264

\bibitem[{{Wijnands} et~al.(2003){Wijnands}, {van der Klis}, {Homan}
  et~al.}]{Wijnands03}
{Wijnands} R., {van der Klis} M., {Homan} J., et~al., 2003, \nat, 424, 44

\bibitem[{{Yin} et~al.(2007){Yin}, {Zhang}, {Zhao} et~al.}]{Yin07}
{Yin} H.X., {Zhang} C.M., {Zhao} Y.H., et~al., Aug. 2007, \aap, 471, 381

\bibitem[{{Zhang} et~al.(1993){Zhang}, {Giles}, {Jahoda} et~al.}]{Zhang93}
{Zhang} W., {Giles} A.B., {Jahoda} K., et~al., 1993, In: Proc. SPIE Vol. 2006,
  p. 324-333, EUV, X-Ray, and Gamma-Ray Instrumentation for Astronomy IV,
  Oswald H. Siegmund; Ed., 324--333

\end{thebibliography}
\end{document}